\documentclass[nobalancelastpage,rmp,showkeys,showpacs,raggedbottom,
  twocolumn,a4paper
]{revtex4-1}
\pdfoutput=1
\usepackage{amsmath}
\usepackage{amssymb}
\usepackage{amsthm}

\newcommand{\Umin}{
    U_\mathrm{min}
    }

\newcommand{\LL}{
  \mathcal{L}
  }

\newcommand{\VV}{
  \mathcal{V}
  }

\renewcommand{\phi}{
  \varphi
}

\newcommand{\toTL}{
  \stackrel{L\to \infty}{\to}
}

\newcommand{\tc}{
  T_c
  }

\newcommand{\cdiff}{
  C^+
  }
\newcommand{\vev}[1]{
  \langle #1 \rangle
  }
\newcommand{\phisp}{
  \phi_\mathrm{sp}
}
\newcommand{\Msp}{
  M_\mathrm{sp}
}

\newcommand{\msp}{
  m_\mathrm{sp}
}
\newcommand{\mf}{
  m_\mathrm{f}
}

\newcommand{\mdiff}{
  \Delta m
}

\newcommand{\mt}{
  \mdiff_\mathrm{t}
  }
\newcommand{\Cin}{
  \mathcal{C}_\text{in}
}

\newcommand{\Cout}{
  \mathcal{C}_\text{out}
}

\newcommand{\bigo}[1]{
  \mathrm{O}(#1)
}
\newcommand{\oo}[1]{
  \mathrm{o}(#1)
}
\newcommand{\rmd}{
  \mathrm{d}
}

\begin{document}
\title{%
Misunderstanding that the Effective Action is Convex under Broken Symmetry
}

\author{\surname{ASANUMA} Nobu-Hiko}
\email{phys.anh@z2.skr.jp}
\homepage{http://z2.skr.jp/phys/}
\affiliation{No research affiliation}

\begin{abstract}
  The widespread belief that the effective action is convex and has a flat bottom under broken global symmetry is shown to be wrong. We show spontaneous symmetry breaking necessarily accompanies non-convexity in the effective action for quantum field theory, or in the free energy for statistical mechanics, and clarify the magnitude of non-convexity. For quantum field theory, it is also  proved that translational invariance breaks spontaneously  when the system is in the non-convex region, and that different vacua of spontaneously broken symmetry cannot be superposed.
  As applications of non-convexity, we study the first-order phase transition which happens at the zero field limit of spontaneously broken symmetry, and we propose a simple model of phase coexistence which obeys the Born rule.
\end{abstract}


\maketitle

\section{Introduction}
The effective action $\Gamma[\vev{\phi(x)}]$ is one of the most fundamental objects in  quantum field theory. Much has been said about it under broken global symmetry.  For simplicity, we here use the language of a field theory with one scalar field $\phi$, symmetric under the operation $\phi \to -\phi$. Let the field has the non-zero vacuum expectation value $|\vev{\phi}| = \phisp$. It is widely believed that:
    ``Because (A) $-\Gamma$ is downward convex, (B) it has a flat bottom for $|\vev{\phi}| < \phisp.$ (C) States in the flat bottom are realized as a linear superposition of the vacua $|\pm \Omega \rangle$, where $\vev{\pm \Omega | \phi(x) | \pm \Omega} = \pm \phisp$.''

We will show, however, that all the points (A), (B) and (C) are wrong.\footnote{
  Almost ubiquitously \textcite{symanzik1970} is cited, presumably intending the attribution of the point (A). However, \citeauthor{symanzik1970} explicitly excludes the ``flat-bottom'' region from the discussion. See its appendix.}
Based on this misunderstanding, it has been thought that there is contradiction, and the resolution has been attempted. Its history is concisely summarized in the recent article by \textcite{Argyres2009.pathIntegral}.  See also \textcite[][sec. 2.2.5]{Delamotte} and \textcite{1126-6708-2007-04-051} for  recent cases where convexity is assumed.

The origin of this fallacy probably comes from the similarity of the effective action with the free energy in statistical mechanics. For example, consider Ising ferromagnet where temperature $T$ is $ < \tc$, the critical temperature, so there is spontaneous magnetization $M = \pm \Msp$, where $M$ is the total magnetization. Let $F(\beta, h)$ be the free energy, where $\beta = 1/T$ and $h$ the external field, and its Legendre transform $C(\beta, M) := F + hM.$ $C$ plays a role similar to the effective potential $\VV (\vev{\phi})$, which is the density of the effective action with uniform vacuum configuration.

In fact, although $F$ is always convex,\footnote{
  Do not confuse it with (exotic) non-convex $F$ which has been of recent interest. For examples, see \textcite{sethna}.
}
$C$ is not for $|M| < \Msp$. It's because it is the region of phase coexistence, and there  the cost of  domain wall formation arises. So the non-convexity  of $C$ is of order $\bigo{L}$ in the 2-dimensional Ising model, where $L$ is the linear extent of the system.

Oddly enough, this fact has already been known for long in the communities of  mathematics oriented statistical mechanics and Monte Carlo simulation.\footnote{
  One of important uses of $C$ in the phase coexistence region is to obtain the interface free energy, for example in Wulff construction. For reviews, see \textcite{Miracle-Sole:2013,Abraham}.
  The interface free energy calculation is an old subject also in Monte Carlo simulation. \cite{PhysRevA.25.1699}
}
Still, the integral understanding of non-convexity is lacking.\footnote{
  For example \textcite[][sec. 1]{Ioffe1994} says to the effect that the $C/V$ is strictly convex and has a flat bottom, but the decay rate of $\exp (-\beta C)$ is of interface order in the 2-dimensional Ising model. This statement is correct, but generality and accuracy are not sufficient.  In a Monte Carlo simulation article, \cite[][sec IV. A]{Nogawa2011} at one hand it is said that there is a maximum in the free energy, while on the other hand the free energy is convex, showing some confusion.
  } As a result, arguments on non-convexity is often too much detailed and technical, obscuring the essence. In the rest areas of physics, non-convexity is barely known, not to mention field theorists, leading to confusion.

As we will show, spontaneous symmetry breaking always accompanies non-convexity, which is never zero. We will further prove that the order of non-convexity is always $\oo{V}$, i.e. strictly smaller than the volume, including the cases of continuous symmetry, in which domain wall cannot form.

Non-convexity is therefore indispensable to understand the correct vacuum in  quantum field theory, or the correct equilibrium in statistical mechanics. We here emphasize that non-convexity itself is so easy that it should be learned  even by undergraduate students. For example Landau's theory of phase transition is  a standard topic in textbooks of statistical physics. It is wrongly explained to be a theory with defect by its non-convexity. We will be able to see how it can be correctly understood.

For quantum field theory, two important consequences of non-convexity will be also shown. One is the impossibility of superpositions like $\lambda |+ \Omega\rangle + (1-\lambda) |- \Omega \rangle$, the point (C) mentioned above. This point has remained ambiguous to date. The other one is the spontaneous breakdown of the translational invariance inside the region of non-convexity.  These two are true for general systems. Although nothing sure can be said except for the Ising model/the quantum theory with one scalar field, the region of non-convexity seems to introduce topological defects naturally.

This paper is organized as follows. First in section \ref{section-lattice} we study the statistical mechanics of lattice systems. To be concrete, we first draw on an exact result on Ising model before obtaining general results. As an application, in section \ref{section-first-order-transition} we will  examine the first-order phase transition at $|M| = \Msp.$ We will find a slight shift of the transition point, and obtain a result more general  than already known one. Precise degree of  non-convexity in general lattice systems is stated in section \ref{section-general-lattice}.

In section \ref{section-qft} we study how the results on lattice models are recast into continuous field theory.
We will see that the conventional notion of linear superposition can be too na\"ive in thermodynamic limit.  In section \ref{section-born}, as another application of the non-convexity of the free energy, we model the phase coexistence of lattice systems, and obtain a simple system which obeys the Born rule. Section 5 gives the conclusion.

\section{Statistical mechanics of lattice systems}
\label{section-lattice}
\subsection{2-dimensional Ising model results}
In this section we study spontaneous symmetry breaking of lattice systems. We include some basic notes so that it will be readable also to students. Rather than stating general results from the beginning, until section \ref{section-first-order-transition} we take up the 2-dimensional square Ising ferromagnetic model. It has its own right, since it is an exceptional case where the explicit dependence of the free energy on the magnetization is known. While it is not as complete as Onsager's solution,  it is remarkable, since in the latter the dependence on the external field is unknown. We think it deserves to be known by every physicist.

We define our model together with relevant symbols as:
\begin{align}
  Z(\beta, h)  & := \sum_{\{\sigma_i  = \pm 1\}} \exp(-\beta E(h)),  \\
  E(h)  & := - J \sum_\mathrm{n.n.} \sigma_i \sigma_j - hM,  \\
  M  & := \sum_i \sigma_i,  \\
  m & := M / V,  \\
  V & := L^2,\\
  \label{F}
  \beta F(\beta, h)  & := - \log Z, \\
  \label{C}
  \begin{split}
  \beta C(\beta, m)  & := \beta(F + hM)\\
  & = -\log \sum_{\{\sum \sigma_i = M\} } \exp(\beta J \sum_\mathrm{n.n.} \sigma_i \sigma_j).
  \end{split}
\end{align}
Let the interaction be nearest-neighbor, and $m$ the per-spin magnetization. When $T < \tc$, the system exhibits the spontaneous magnetization $m = \pm \msp(T)$ at zero field limit $h=\pm0.$ $C(\beta, m)$ is the free energy when $(\beta, m)$ are the variables that specify the state of the system, so $F(h=\pm 0) = C(m = \pm \msp).$ We will often present statements and equations which are exactly correct and/or meaningful only in thermodynamic limit  $L \to \infty$, but readers will have no difficulty understanding them.

The case where $|m| < \msp$ is of our interest. Its entire region corresponds to $h=0$. (Consider vapor-liquid coexistence of water.) As we will see, it is the region of phase coexistence, but we call it more generally as the ``region singular with respect to the external field'', or RSEF.

We want to write down $C$. To this end, notice $Z(h=0) = \sum_m \exp(-\beta C(m)),$ so $P(m') := \exp(-\beta C(m')) / Z$ is the probability that $m$ takes the value $m'$ when $(T, h = 0)$ are specified. This is exactly the inverse of deriving grand canonical ensemble from canonical ensemble. The explicit formula of $P$ is obtained for the 2-dimensional Ising model for free and periodic boundaries in thermodynamic limit, at least for low enough temperature. \cite{shlosman1989} To rewrite it for $C$, first let
\begin{multline}
  \label{Cin}
  \Cin(m) := C(\msp) + \\
  \quad \begin{cases}
    aLT \sqrt{\msp - |m|} & \text{for } \mf < |m| < \msp, \\
    aLT \sqrt{\msp - \mf\phantom{|}} = \mathrm{const.} & \text{for } |m| < \mf
  \end{cases}
\end{multline}
where $\mf > 0$ and $a > 0$ are finite constants dependent on $\beta$ and the boundary condition. Then for  $|m| < \msp$,
\begin{align}
  \label{Ceq}
  \begin{split}
  C(m) =& \Cin(m),  \\
  \cdiff(m) := & C(m) - C(\msp).
  \end{split}
\end{align}
We have expressed $C$ in two steps because $\Cin$ is defined only in the RSEF, while $C$ is defined in the entire $(T, m)$ domain. Consider $\Cin$ as a pure mathematical function, while $C$ the free energy, a physical entity.

It is now obvious that the claims (A), the convexity of $C$, and (B), the ``flat bottom'' are illusions. $C$ is \emph{not} always downward convex. We here clarify the flaw in the proof of $C$'s convexity: $F$ is convex inside any \emph{finite} interval of $h$, and this fact has to be used to prove  $C$'s convexity. But  the RSEF corresponds to one point $h=0$. Thus $F$'s convexity does not imply that $C$ is always convex.

An intuitive interpretation of $\cdiff$ is desirable. \cite{shlosman1989}
To avoid unnecessary complication, we take the free boundary. When $ |m| < \mf$ (``flat region'' hereafter), the system consists of two domains, the plus-spin rich phase and minus- rich one, separated by a straight wall, running parallel to the two system edges. (Although 2-dimensional, we use the word domain \emph{wall}.) As long as $m$ stays in the flat region, the wall can shift but remains straight, and $C$ does not change. When $\mf < |m| < \msp$, the smaller domain favors to be a droplet\footnote{
  The word ``droplet'' is a standard nomenclature. Here the droplet is sole and macroscopic, and in equilibrium. It is also used in non-equilibrium nucleation in usual first-order phase transitions, where there are many droplets, and the growth of initially microscopic droplets to macroscopic size is studied.
}
at a corner to lower the free energy.
We immediately recognize that (i) the RSEF is a phase coexistence region, (ii) $L\sqrt{\msp - |m|}$ is of the order of the wall length $l$ and $\cdiff$ can be written $= sl$, where $s$ is the interface free energy per unit length, and (iii) the flatness of $C$  means the wall's thickness is finite, or the wall is ``sharp'', so does not interact with the boundaries. (When the boundary is periodic, there are two walls, and they don't interact with each other.) For general boundaries the shape of the domain is determined by ``Wulff construction'' For reviews, see \textcite{Miracle-Sole:2013,Abraham}.

There is a phase transition at $|m| = \mf$ which is sometimes called droplet-strip transition.
It is obviously first-order; both domains discontinuously change their geometries, and both sides of the transition point have the metastable branch.

\subsection{Normalization---first encounter}
To understand non-convexity correctly, we have to be careful about the normalization of variables. This is the key to our entire argument.

All extensive variables scales as $V$, or put differently, are $\bigo{V}.$ So it is tempting to consider ``free energy per spin'' $c(m) := C(m)/ V.$ Aside from its finiteness, $c$ has another advantage over $C$ that it's differentiable with $m$: $C$ was originally defined for $M$, a discrete variable, but we would like to define $\partial C / \partial M$ and to expect it to be $=h$. There is no problem by defining it as $(C(M + n) - C(M)) / n$ $(n \in \mathbb{Z})$, since $\to \partial C/ V \partial m$ as $L\to\infty.$ This means $\partial c /\partial m = h$. Then $c(m) = c(\msp)$ throughout the RSEF since it corresponds to $h=0$. It is of course consistent with the explicit formula (\ref{Cin}) of $\Cin$. So $c(m)$ is indeed differentiable and globally convex.

Notice at the same time $\cdiff(m) \to + \infty$ when $|m| < \msp$. It has to be so, because $P(m) \propto \exp(-\beta C(m)).$ In words, it is necessary in order for spontaneous symmetry breaking to happen.

The conclusion of this short section is: $c(m)=c(\msp)$ inside the RSEF means $c$ does not contain the information on the RSEF. Sticking to $c$'s convexity is the same as considering $F(\beta, h)$ and neglecting the RSEF. Correct normalization is to consider $\cdiff / L.$

Two pedagogical notes are in order: (i) By the ``Maxwell construction'' the flat-bottomed $C$ can be obtained. It is equivalent to divide the system into the domains of different phases, and to ignore the domain wall free energy. It will not be much problematic, or rather better e.g. when considering bulk equilibrium in chemical reaction. (ii) The free energy is equivalent to the partition function, in the sense that in the free energy all information of the system is contained. (Prosaically plug $\exp(-\beta F)$ in place of $Z$, and you're done.) Unnecessary approximation of the free energy throws away some information. In field theory, the counterparts of free energies are generating functionals of connected diagrams.

\subsection{Phase transition at $|m| = \msp$}
\label{section-first-order-transition}
Next we have a close look at the phase transition at $\mdiff := \msp - |m| = 0.$ It is an application and not necessary to understand non-convexity, so uninterested readers can skip this section.

Because we are using the extensive variable $M$ (in fact $m$, a normalized one), the transition looks continuous. More precisely, it is clearly first-order when approached from the outside of the RSEF, as $|m| \searrow \msp,$ by not showing any sign of transition beforehand, and the transition happens suddenly at $\mdiff = 0$. After the system passes the transition point, it can go to the metastable branch, too. On the other hand, inside the RSEF there is the relation $\cdiff \propto \mdiff^{1/2}$ which looks like scaling laws, just like continuous transitions. This is partly expected, because phase coexistence terminates at the point $\mdiff = 0,$ or put differently cannot extend over that point. So there must be a singularity, and by definition, $\cdiff = 0$ at $\mdiff = 0.$ There cannot be latent heat and metastability after passing the transition point...

In reality this transition is \emph{not} continuous and accompanies discontinuity. First let us write:
\begin{equation}
  C(m) = \Cout(m) \quad \text{for} \quad m > \msp.
\end{equation}
Then $\Cout$  is regular at $m = \msp$:
\begin{equation}
  \label{Dreg}
  \Cout(m) := C(\msp) + L^2(c_2 \mdiff^2 + c_3 \mdiff^3 + ...),
\end{equation}
where $c_2 = (\partial m/\partial h)^{-1}|_{h=0}$, and $c_2, c_3 = \bigo{1}.$
Notice that $\Cout$ is defined inside of RSEF too, so  compare it with $\Cin$. Then $aLT \mdiff^{1/2} > c_2 L^2 \mdiff^2$, i.e. $\Cin > \Cout$  for small enough $\mdiff >0.$ Therefore the true transition happens at $|\mdiff| = \mt,$ slightly inside the RSEF, where $\mt = aTc^{-1}_2 L^{-2/3},$ or at $|M|= M_\mathrm{t}$ where $\Msp - M_\mathrm{t} \propto L^{4/3}$. Although $\mt \to 0$ in thermodynamic limit, this shift is observable since $\Msp - M_\mathrm{t} \to \infty.$  So at the transition point the big one droplet appears/disappears all of a sudden, and this discontinuity is inevitable even if the \emph{extensive} variable $M$ is utilized. Latent heat must be there, too. In order to avoid discontinuity completely, $E$ or $S$, the entropy, has to be used instead of $T$. Also notice that $\cdiff \propto \mdiff^{1/2}$ represents a (small) metastable branch for $0 < \mdiff < \mt.$

The above shift of the transition point was first proved rigorously by \textcite{Biskup:EPL:2002,Biskup:CMP:2003}. They also showed it by a physical argument which is applicable to general vapor-liquid like transitions, which shares much with our argument. Although they treat it more deeply, ours can be applicable to more general systems, and in fact we will do so in the next section. Similar shift is observed for other systems, for example in a Monte Carlo study of Potts model. \cite{Nogawa2011} The shift is an important factor for simulations because it is a finite size effect.

Another pedagogical remark: If $\Cout$ is regular, why does not it extend into the RSEF, like analytic continuation? It's because $\Cin < \Cout$ in the RSEF.\footnote{
  Equation \eqref{C} can be rewritten as $-\beta C(m) = \log \sum_\nu \exp(-\beta C_\nu(m))$, where $C_\nu(m) := E_\nu(m) - TS_\nu(m)$, $E_\nu(m)$ is the energy of the $\nu$-th level with $m$, and $\exp(S_\nu(m))$ is the number of states in the $\nu$-th level with $m.$ Because $C = \bigo{V},$ only the vicinity of the $C_\nu$'s minimum can contribute in thermodynamic limit.
}  \footnote{
  More note on analyticity: In general free energy is a patchwork of regular functions. For example, consider the system with a sole ground state at $E = 0$, and $\exp(aN)$-fold excited states at $E = \epsilon N$. Then $Z = 1 + \exp((a-\epsilon \beta) N).$ This $Z$ is analytic for finite $N$, but in ``thermodynamic limit'' $N \to \infty$,
  \begin{equation*}
    F = \begin{cases}
      F_{-} := 0 & \text{for} \quad aT < \epsilon,\\
      F_{+} := (\epsilon - aT)N  & \text{for} \quad aT > \epsilon.
    \end{cases}
  \end{equation*}
  Notice both $F_{\pm}$ are analytic for all $T$, but the smaller one is ``chosen''. They get cut and sewn together at $aT = \epsilon$. If a singularity of $F$ has nothing to do with that of $F_{\pm}$, then it is the first-order transition point.
  }
 Or $\Cin$ emerges and masks $\Cout$, so to say. Still, $\Cout$ naturally extends into the RSEF, and describes the metastable branch---it is a realm of non-equilibrium theory and not part of the Boltzmann's statistical mechanics. We omit this subject.

\subsection{Non-convexity and spontaneous symmetry breaking; general lattice systems}
\label{section-general-lattice}
We consider spontaneous symmetry breaking of general lattice theories. Although there may be multiple fields and conjugate extensive variables, we suppress labels to distinguish them, and continue to use the same symbols with the former ones; meaning must be clear. Let the dimensionality be $d$, $V=L^d$, and the symmetry is spontaneously broken at $h=0$.  When the symmetry is continuous,  domain walls cannot form,
and so the RSEF is not a region of phase coexistence.

The precise magnitude of non-convexity is as follows:
\theoremstyle{definition} \newtheorem*{Theorem}{Theorem}
\begin{Theorem}[Magnitude of non-convexity]
  Suppose there is spontaneous symmetry breaking, being $|m| = \msp$ under zero external field. For $|m| < \msp$, (i) $C(m) - C(\msp) \toTL + \infty$, because it is necessary for spontaneous symmetry breaking to happen. (ii) $C(m) - C(\msp) = \oo{V}$, because it is equivalent to $\partial c/\partial m = h = 0$.
\end{Theorem}

The following two points should be common, but other  cases are not denied completely: (i) If we write $C = \Cout$ outside of the RSEF, $\Cout$ is regular at $m = \msp$. (ii) Inside the RSEF, $C = \Cin$ and $\Cin$ has a singularity at $m = \msp$.

We then study the shift of transition point from $\Msp$, generalizing the previous section. Let $\Cout(m) = C(\msp) + V(c_2 \mdiff^2 + c_3 \mdiff^3 + ...)$, and $C(m) = \Cout(m)$  outside the RSEF. Let $\epsilon > 0 $ be such that $0 < \cdiff / L^{d-\epsilon} < \infty$ inside the RSEF in thermodynamic limit. We now assume a scenario where $\cdiff$'s dependence on $L^{d - \epsilon}$ comes exclusively from $M$'s power inside the RSEF, i.e.
\begin{equation}
  \cdiff \propto L^{d - \epsilon} \propto (\Msp - M)^{(d-\epsilon)/d} = (L^d \mdiff)^{1-\epsilon/d}.
\end{equation}
In this case, $\Cin$ is \emph{upward} convex and $\Cin$ is bigger than $\Cout$  for $1 \gg |\mdiff| > 0$. The transition point then lies always inside the RSEF at $|\mdiff| = \mt \propto L^{-d\epsilon/(d + \epsilon)} \toTL 0$ or at $|M| = M_t $ where  $\Msp - M_t \propto L^{d^2/(d + \epsilon)} \toTL \infty.$ This conclusion is slightly more general than \textcite{Biskup:EPL:2002}, by relating the order of $\Cin$'s singularity with the shift of the transition point. When there is a domain wall, $\epsilon$ is likely to be $ = 1$, but can be other values for continuous symmetries.

\subsection{General phase equilibria}
We also consider non-lattice systems quickly. As we already stated, in general phase equilibria like chemical one, the free energy in extensive variables is not convex by the order of interface. The interface free energy has to be positive; otherwise, bubbles would form and phases would mix. Pedagogical: When there is no  spontaneous symmetry breakdown, there are  non-zero external fields, so the free energy per volume under phase coexistence is linear, but not constant.

\section{Quantum field theory}
\label{section-qft}
\subsection{Quantum theory of one scalar field}
It should  already be clear that in (continuous) quantum field theory under spontaneous broken symmetry, the  effective action $\Gamma[\vev{\phi(x)}]$ is not convex. However conversion from a lattice theory to quantum field theory has some non-trivial points, if straightforward. Again, normalization is the point. To set up a necessary language, we start  this section by studying a theory with one scalar field.

We first remember how to obtain the $\phi^4$-theory from the Ising model of $d$-dimension:\footnote{This is a standard procedure. See for example \textcite[][Chap. 4]{Altland-Simons}.
} Convert the spin variables $\{\sigma_i\}$ to real-valued variables $\{\phi_i\}$ by Hubbard-Stratonovich transform and take some more steps to clean up. Notice the condition $\sum \sigma_i = M$ is mapped to the equation
\begin{equation}
  \label{op-id}
  \sum \phi_i = M,
\end{equation}
since $T \partial \log Z / \partial h_i = \langle \sigma_i \rangle= \langle \phi_i \rangle$ (with appropriate normalization), where brackets are vacuum expectation value and $h_i$  is the per-site external field. Equation \eqref{op-id} is an operator identity, because $Z$ is the generating functional. Finally Fourier-expand $\phi_i$ and retain only low-frequency modes. Then the constraint \eqref{op-id} becomes $\tilde{M} := \int \rmd^d x \, \phi$.

We have to deal with the obstacle that $m$ was specified in the lattice system. This obviously corresponds to non-uniform vacuum. We continue to use $L$ to mean the system's linear extent. To let the symmetry  break spontaneously we have to take $L \to \infty$. Two cases are possible: (i) If you keep the domain wall near the origin, then the wall gets stretched into a straight line, with the boundary conditions say $\langle \phi \rangle \to \pm \phi_\mathrm{sp}$ as $x^1 \to \pm \infty,$ and the original condition of $m$ has vanished. Or (ii) if the domain wall is sent infinitely far as $L \to \infty$,  we get the theory $\langle \phi \rangle = \pm \phi_\mathrm{sp},$ namely the theory with uniform vacuum, even though we started from the lattice theory inside the RSEF. This is not intriguing. In both cases, we assumed that we can neglect the constraint about $\int dx \phi(x)$, which is nonlocal. This must be no problem insofar as there are only finite excitations. Variant models in which the system edge is there, e.g. where $y$ is limited to $\ge 0,$ are also possible.\footnote{
In this case, renormalization at the system boundary is necessary. For reviews on surface/interface renormalization see e.g. \textcite{Diehl:domb-v10,Jasnow:domb-v10}
}

The condition on $M$ in the lattice theory seems to have disappeared in the field theory. It is because the purposes of the theories are different---in field theories global quantities like total magnetization are not of much concern, and correlation functions are of prime importance. When the non-convexity of the Landau theory of phase transition is concerned for instance, you can focus on $\Gamma$ with proper normalization. The situation is actually the same in lattice theories; $M$ is irrelevant to calculate correlation functions in thermodynamic limit.

The vacuum with a wall is usually considered to be a topological condition. In the present discussion it was naturally introduced as the normalization condition which was originally there in the lattice theory.

We now explicitly demonstrate the non-convexity. We take Minkowski spacetime, but it applies to Euclidean cases too with obvious modification. Let the space dimensionality be $d-1$, $T :=  \int dt$, $V := T \int \rmd^{d-1} x$, and the Lagrangian density $\LL$ have the potential $U$:
\begin{equation}
  \LL [J] := \frac{1}{2}  ( \partial_\mu \phi)^2 - U(\phi) + \phi J,
\end{equation}
where $U(\phi) = U(-\phi)$.
 We define $\Gamma$ as:
\begin{align}
  \mathrm{e}^{i \Gamma [ \vev{\phi}_J ]} &:= \nonumber \\
  & \int \mathcal{D} \phi \exp
  \left (i \int \rmd^d\, x ( \LL [ J ] - \vev{\phi(x)}_J J(x)) \right),\\
  \vev{\phi(x^\prime)}_J & := -i \frac{\delta}{\delta J(x^\prime)} \log \int \mathcal{D} \phi \exp(i \int \rmd^d\, x \LL [ J ]),
\end{align}
so that $- \Gamma$ takes the minimum for $|\vev{\phi(x)}| = \phisp := \vev{\phi}_{+ 0}$. $\Gamma$ is related to the vacuum energy $E_\mathrm{vac}$  by:
\begin{equation}
  - \Gamma = TE_\mathrm{vac}.
\end{equation}
(\citet{symanzik1970}; for an introduction, see \textcite[][Sec 16.3]{WeinbergQFT:2})

We now calculate the vacuum energy when a wall is present at the classical level.\footnote{
  See \textcite{Bogomolny:1975de}. For an introduction see also \textcite[sec. 23.1]{WeinbergQFT:2}. Under spontaneous breakdown of spacetime symmetry, counting the number of Goldstone bosons is tricky. See \textcite{PhysRevLett.88.101602}.
} Let $U$ take the absolute minimum $\Umin$ at $\pm \phi_\mathrm{sp}.$
We impose the boundary condition $\phi(x) \to \pm \phi_\mathrm{sp}$ as $x^1 \to \pm \infty,$ and uniform in time and other spatial directions.
(The argument when the boundary condition is spatially uniform and temporarily non-uniform is essentially the same.)
Then the Hamiltonian $H$ has the minimum value, or the vacuum energy
\begin{equation}
  \label{Evac}
E_\mathrm{vac} := L^{d-2} \int^{\phi_\mathrm{sp}}_{-\phi_\mathrm{sp}} d\xi \sqrt{2U(\xi)} + L^{d-1} \Umin,
\end{equation}
with the vacuum field configuration which satisfies
\begin{equation}
  \frac{\partial}{\partial x^1} \phi = \sqrt{2U(\phi)}.
\end{equation}
The first term of the right hand side of equation \eqref{Evac} is the non-convex part. Since $\Umin$ is $\bigo{1}$, notice that $\Gamma = \bigo{V}$, although $\Gamma$ itself is dimensionless. What is important here is a mere dimensional analysis and quantum correction should not alter the conclusion.

\subsection{General field theory; spontaneous breakdown of translational invariance}
\label{section-TL}
As we saw in the previous section, $\Gamma$ is $\bigo{V}$, but the non-convexity is of order $\oo{V}$. This is necessarily so for any quantum field theory with spontaneous symmetry breakdown, as we saw in the case of lattice theories.

Now we prove that  the translational invariance is spontaneously broken inside the region singular with the external field, namely the region where $\Gamma$ is non-convex. We use the notation of the theory with one scalar field, but it obviously applies to any symmetry.

If  the vacuum configuration $\vev{\phi}$ inside the non-convex region were temporarily and spatially uniform, then $- \Gamma[\vev{\phi}] > -\Gamma[\phisp]$, by definition of the spontaneously symmetry breakdown. But then the left hand side would be bigger than the right hand side by the order $V$. This is a contradiction. $\square$

This can be paraphrased as: By controlling $\vev{\phi}$, spontaneous symmetry breakdown as $|\vev{\phi}| = \phisp$ is lost, but it has changed its guise as the spontaneous breakdown of the translational invariance. We failed to prove this result for statistical mechanics.

\subsection{(Seeming) breakdown of linearity in thermodynamic limit}
It is clear (was already from the discussion of lattice systems) that the statement (C), the states in the non-convex region are linear superpositions of the pure-phase vacua, is wrong.

There is one important corollary: different vacua of spontaneously broken symmetry cannot be superposed. In thermodynamic limit, such thing can happen. It is because: (i) If it were possible to add the two vacua $|\pm \Omega \rangle,$ then spontaneous symmetry breaking couldn't happen, by allowing intermediate states. (ii) $|\pm \Omega\rangle$ belong to two different theories, described by  to two different Lagrangians $\mathcal{L}[\pm 0]$.  These two explanations would sound somewhat heuristic without our result, but we can now be confident.

With the last explanation (ii) at hand, a more accurate statement will be: You can not even think of adding $|\pm \Omega \rangle$, which belong to different theories.

Of course, in spontaneous symmetry breaking only one vacuum is believed to be chosen. Still, this point has not met with a firm, clear consensus to date; \textcite[Sec 16.3]{WeinbergQFT:2} claims the superposition of different vacua. See also \textcite{Argyres2009.pathIntegral} and references therein for the confusion on this point. We have finally settled it.

In fact, denial of superposition of $|\pm \Omega\rangle$ to that effect was already indicated by \textcite[][Chap 5]{aspects-of-symmetry} and by \textcite[][Sec 19.1]{WeinbergQFT:2} under different contexts, each of which is physically insightful, if not rigorous. The former discusses the instability of large systems under small perturbation, and the latter the requirement of cluster decomposition.

Once it is shown, it may seem a trivial matter, but it will have serious influence over the discussion of quantum measurement. There are manifold ``interpretations'', but usually linearity is considered to be never broken. That's why some people proposed theories which can be categorized as ``dynamical reduction models''---those who work in this direction believe that the wave function indeed collapses, and because the Schr\"odinger equation is strictly linear, they modify it to induce collapses. Neither do proponents of decoherence programs think that linearity can be violated. For reviews, see \textcite{RevModPhys.85.471, RevModPhys.76.1267, 1.1356698, Bassi2003257}.

Our result indicates that pure quantum theory  intrinsically has a room for  non-linearity---to be more accurate, what can be superposed is not self-evident in thermodynamic limit. Arguments based on systems with finite degrees of freedom are invalid.

\section{A model of random walk of domain wall}
\label{section-born}
\subsection{Definition and result}
As another application of the non-convex free energy of the 2-dimensional Ising model given in equation \eqref{Cin}, we consider a simple, abstract model which simulates the fluctuation of a rigid domain wall.

Suppose the ``wall'' is initially in the flat region (i.e. $|M| < M_\textrm{f}$) and then the total magnetization is allowed to variate ``quasi-statically''. Consider a one dimensional lattice whose points are denoted by $x \in \{-M_\mathrm{f}, -M_\mathrm{f} + 1, ..., M_\mathrm{f}\}, M_\mathrm{f} \in \mathbb{N}$.  The wall randomly walks on this lattice. Call the regions left and right to the wall as ``spin-plus phase'' and ``-minus'' one, respectively. At each time step the wall moves to one of the neighbor points with equal probabilities 1/2. When the wall reaches an end point $x = \pm M_\mathrm{f}$ so getting out of the flat region, then the wall rolls down the slope of the free energy irreversibly, and the system ends up in a uniform state.

This simplest model is unique in the following point: If the wall is initially at $x$, or equivalently the plus region shares $v := (x + M_\mathrm{f}) / 2M_\mathrm{f}$ part of the system's whole volume $2M_\mathrm{f}$, then  the probability that the final state is the plus phase is $v$, which is easy to prove.

This immediately allows to be made more abstract and more general: Suppose (i) each spin can take $q$ states, (ii) there can be arbitrary number of walls and (iii) when any of two walls meet, or a wall reaches the system edge, they are annihilated. The boundary can be periodic, but it does not matter. Again, the probability that $i$-th state domain is the only survivor is proportional to its volume. To prove it, notice that the state can be expressed as a point on the lattice on a surface of $q$-simplex, where $x_i \in \{0, 1, ..., N\}$ and $ \sum x_i = N.$ The state hops to one adjacent point with equal probabilities. Once one coordinate $x_i$ gets 0, it will be opted out, and the system will be in $(q-1)$-simplex, and so on.

The above models satisfy two important conditions of quantum measurement, the irreversible collapse of the state and the Born rule, and we \emph{imagined} that (i) the time development is deterministic, (ii) randomness is brought about by noise,  and (iii) large degrees of freedom (thermodynamic limit) is necessary.

\subsection{Discussion}
Probably a pure mathematical model which obeys the Born rule and is equivalent to ours has already been reported; they are the simplest models of random walk, after all. (Possibly underneath any theory which obeys the Born rule our model might be hidden.) But  by the present models we would like to encourage to take the Copenhagen interpretation more seriously: In finite systems, time evolution is unitary, but in thermodynamic limit, it is not always so. Admittedly we cannot hint at anything on the model's relation to quantum mechanics.

\section{Conclusion}
We studied lattice statistical mechanics and quantum field theory under spontaneously broken symmetry, and found that the free energy and the effective action are not convex. The order of non-convexity is found to be strictly smaller than the system's volume, but it has to diverge in thermodynamic limit. For lattice systems, the first-order transition at $|m| = \msp$ was studied; there is a slight shift of the transition point, and we showed it in a way which may be applicable when the symmetry is continuous. For quantum field theory, we proved that the translational invariance breaks in the non-convex region.  We proved that different vacua of spontaneously broken symmetry can not be added, which has been ambiguous. We also modeled the phase coexistence of the Ising model, and obtained a simple model which obeys the Born rule.

\bibliographystyle{apsrmp4-1}
\bibliography{bib/phys.bib}

\begin{thebibliography}{25}%
\makeatletter
\providecommand \@ifxundefined [1]{%
 \@ifx{#1\undefined}
}%
\providecommand \@ifnum [1]{%
 \ifnum #1\expandafter \@firstoftwo
 \else \expandafter \@secondoftwo
 \fi
}%
\providecommand \@ifx [1]{%
 \ifx #1\expandafter \@firstoftwo
 \else \expandafter \@secondoftwo
 \fi
}%
\providecommand \natexlab [1]{#1}%
\providecommand \enquote  [1]{``#1''}%
\providecommand \bibnamefont  [1]{#1}%
\providecommand \bibfnamefont [1]{#1}%
\providecommand \citenamefont [1]{#1}%
\providecommand \href@noop [0]{\@secondoftwo}%
\providecommand \href [0]{\begingroup \@sanitize@url \@href}%
\providecommand \@href[1]{\@@startlink{#1}\@@href}%
\providecommand \@@href[1]{\endgroup#1\@@endlink}%
\providecommand \@sanitize@url [0]{\catcode `\\12\catcode `\$12\catcode
  `\&12\catcode `\#12\catcode `\^12\catcode `\_12\catcode `\%12\relax}%
\providecommand \@@startlink[1]{}%
\providecommand \@@endlink[0]{}%
\providecommand \url  [0]{\begingroup\@sanitize@url \@url }%
\providecommand \@url [1]{\endgroup\@href {#1}{\urlprefix }}%
\providecommand \urlprefix  [0]{URL }%
\providecommand \Eprint [0]{\href }%
\providecommand \doibase [0]{http://dx.doi.org/}%
\providecommand \selectlanguage [0]{\@gobble}%
\providecommand \bibinfo  [0]{\@secondoftwo}%
\providecommand \bibfield  [0]{\@secondoftwo}%
\providecommand \translation [1]{[#1]}%
\providecommand \BibitemOpen [0]{}%
\providecommand \bibitemStop [0]{}%
\providecommand \bibitemNoStop [0]{.\EOS\space}%
\providecommand \EOS [0]{\spacefactor3000\relax}%
\providecommand \BibitemShut  [1]{\csname bibitem#1\endcsname}%
\let\auto@bib@innerbib\@empty
\bibitem [{\citenamefont {Abraham}(1986)}]{Abraham}%
  \BibitemOpen
  \bibfield  {author} {\bibinfo {author} {\bibnamefont {Abraham}, \bibfnamefont
  {D.~B.}}} (\bibinfo {year} {1986}),\ in\  \cite{DombLebowitz:10},\
  Chap.~\bibinfo {chapter} {1}\BibitemShut {NoStop}%
\bibitem [{\citenamefont {Altland}\ and\ \citenamefont
  {Simons}(2010)}]{Altland-Simons}%
  \BibitemOpen
  \bibfield  {author} {\bibinfo {author} {\bibnamefont {Altland}, \bibfnamefont
  {A.}}, \ and\ \bibinfo {author} {\bibfnamefont {B.}~\bibnamefont {Simons}}}
  (\bibinfo {year} {2010}),\ \href@noop {} {\emph {\bibinfo {title} {Condensed
  Matter Field Theory}}},\ \bibinfo {edition} {2nd}\ ed.\ (\bibinfo
  {publisher} {Cambridge University Press},\ \bibinfo {address}
  {Cambridge})\BibitemShut {NoStop}%
\bibitem [{\citenamefont {Argyres}\ \emph {et~al.}(2009)\citenamefont
  {Argyres}, \citenamefont {van Kessel},\ and\ \citenamefont
  {Kleiss}}]{Argyres2009.pathIntegral}%
  \BibitemOpen
  \bibfield  {author} {\bibinfo {author} {\bibnamefont {Argyres}, \bibfnamefont
  {E.~N.}}, \bibinfo {author} {\bibfnamefont {M.~T.~M.}\ \bibnamefont {van
  Kessel}}, \ and\ \bibinfo {author} {\bibfnamefont {R.~H.~P.}\ \bibnamefont
  {Kleiss}}} (\bibinfo {year} {2009}),\ \href {\doibase
  10.1140/epjc/s10052-009-1131-y} {\bibfield  {journal} {\bibinfo  {journal}
  {The European Physical Journal C}\ }\textbf {\bibinfo {volume}
  {64}}~(\bibinfo {number} {2}),\ \bibinfo {pages} {319}}\BibitemShut {NoStop}%
\bibitem [{\citenamefont {Bassi}\ and\ \citenamefont
  {Ghirardi}(2003)}]{Bassi2003257}%
  \BibitemOpen
  \bibfield  {author} {\bibinfo {author} {\bibnamefont {Bassi}, \bibfnamefont
  {A.}}, \ and\ \bibinfo {author} {\bibfnamefont {G.}~\bibnamefont {Ghirardi}}}
  (\bibinfo {year} {2003}),\ \href {\doibase
  http://dx.doi.org/10.1016/S0370-1573(03)00103-0} {\bibfield  {journal}
  {\bibinfo  {journal} {Phys. Rep.}\ }\textbf {\bibinfo {volume}
  {379}}~(\bibinfo {number} {5–6}),\ \bibinfo {pages} {257 }}\BibitemShut
  {NoStop}%
\bibitem [{\citenamefont {Bassi}\ \emph {et~al.}(2013)\citenamefont {Bassi},
  \citenamefont {Lochan}, \citenamefont {Satin}, \citenamefont {Singh},\ and\
  \citenamefont {Ulbricht}}]{RevModPhys.85.471}%
  \BibitemOpen
  \bibfield  {author} {\bibinfo {author} {\bibnamefont {Bassi}, \bibfnamefont
  {A.}}, \bibinfo {author} {\bibfnamefont {K.}~\bibnamefont {Lochan}}, \bibinfo
  {author} {\bibfnamefont {S.}~\bibnamefont {Satin}}, \bibinfo {author}
  {\bibfnamefont {T.~P.}\ \bibnamefont {Singh}}, \ and\ \bibinfo {author}
  {\bibfnamefont {H.}~\bibnamefont {Ulbricht}}} (\bibinfo {year} {2013}),\
  \href {\doibase 10.1103/RevModPhys.85.471} {\bibfield  {journal} {\bibinfo
  {journal} {Rev. Mod. Phys.}\ }\textbf {\bibinfo {volume} {85}},\ \bibinfo
  {pages} {471}}\BibitemShut {NoStop}%
\bibitem [{\citenamefont {Binder}(1982)}]{PhysRevA.25.1699}%
  \BibitemOpen
  \bibfield  {author} {\bibinfo {author} {\bibnamefont {Binder}, \bibfnamefont
  {K.}}} (\bibinfo {year} {1982}),\ \href {\doibase 10.1103/PhysRevA.25.1699}
  {\bibfield  {journal} {\bibinfo  {journal} {Phys. Rev. A}\ }\textbf {\bibinfo
  {volume} {25}},\ \bibinfo {pages} {1699}}\BibitemShut {NoStop}%
\bibitem [{\citenamefont {Biskup}\ \emph {et~al.}(2002)\citenamefont {Biskup},
  \citenamefont {Chayes},\ and\ \citenamefont {Koteck{\'y}}}]{Biskup:EPL:2002}%
  \BibitemOpen
  \bibfield  {author} {\bibinfo {author} {\bibnamefont {Biskup}, \bibfnamefont
  {M.}}, \bibinfo {author} {\bibfnamefont {L.}~\bibnamefont {Chayes}}, \ and\
  \bibinfo {author} {\bibfnamefont {R.}~\bibnamefont {Koteck{\'y}}}} (\bibinfo
  {year} {2002}),\ \href@noop {} {\bibfield  {journal} {\bibinfo  {journal}
  {EPL (Europhysics Letters)}\ }\textbf {\bibinfo {volume} {60}}~(\bibinfo
  {number} {1}),\ \bibinfo {pages} {21}}\BibitemShut {NoStop}%
\bibitem [{\citenamefont {Biskup}\ \emph {et~al.}(2003)\citenamefont {Biskup},
  \citenamefont {Chayes},\ and\ \citenamefont {Koteck{\'y}}}]{Biskup:CMP:2003}%
  \BibitemOpen
  \bibfield  {author} {\bibinfo {author} {\bibnamefont {Biskup}, \bibfnamefont
  {M.}}, \bibinfo {author} {\bibfnamefont {L.}~\bibnamefont {Chayes}}, \ and\
  \bibinfo {author} {\bibfnamefont {R.}~\bibnamefont {Koteck{\'y}}}} (\bibinfo
  {year} {2003}),\ \href {\doibase 10.1007/s00220-003-0946-x} {\bibfield
  {journal} {\bibinfo  {journal} {Comm. Math. Phys.}\ }\textbf {\bibinfo
  {volume} {242}}~(\bibinfo {number} {1}),\ \bibinfo {pages} {137}}\BibitemShut
  {NoStop}%
\bibitem [{\citenamefont {Bogomolny}(1976)}]{Bogomolny:1975de}%
  \BibitemOpen
  \bibfield  {author} {\bibinfo {author} {\bibnamefont {Bogomolny},
  \bibfnamefont {E.~B.}}} (\bibinfo {year} {1976}),\ \href@noop {} {\bibfield
  {journal} {\bibinfo  {journal} {Sov. J. Nucl. Phys.}\ }\textbf {\bibinfo
  {volume} {24}},\ \bibinfo {pages} {449}},\ \bibinfo {note} {[Yad. Fiz. 24,
  861 (1976)]}\BibitemShut {NoStop}%
\bibitem [{\citenamefont {Coleman}(1985)}]{aspects-of-symmetry}%
  \BibitemOpen
  \bibfield  {author} {\bibinfo {author} {\bibnamefont {Coleman}, \bibfnamefont
  {S.}}} (\bibinfo {year} {1985}),\ \href@noop {} {\emph {\bibinfo {title}
  {Aspects of Symmetry : Selected Erice Lectures of Sidney Coleman}}}\
  (\bibinfo  {publisher} {Cambridge University Press},\ \bibinfo {address}
  {Cambridge})\BibitemShut {NoStop}%
\bibitem [{\citenamefont {Delamotte}(2012)}]{Delamotte}%
  \BibitemOpen
  \bibfield  {author} {\bibinfo {author} {\bibnamefont {Delamotte},
  \bibfnamefont {B.}}} (\bibinfo {year} {2012}),\ in\ \href {\doibase
  10.1007/978-3-642-27320-9_2} {\emph {\bibinfo {title} {Renormalization
  Group and Effective Field Theory Approaches to Many-Body Systems}}},\
  \bibinfo {editor} {edited by\ \bibinfo {editor} {\bibfnamefont
  {A.}~\bibnamefont {Schwenk}}\ and\ \bibinfo {editor} {\bibfnamefont
  {J.}~\bibnamefont {Polonyi}}}\ (\bibinfo  {publisher} {Springer},\ \bibinfo
  {address} {Berlin})\BibitemShut {NoStop}%
\bibitem [{\citenamefont {Diehl}(1986)}]{Diehl:domb-v10}%
  \BibitemOpen
  \bibfield  {author} {\bibinfo {author} {\bibnamefont {Diehl}, \bibfnamefont
  {H.~W.}}} (\bibinfo {year} {1986}),\ in\  \cite{DombLebowitz:10},\
  Chap.~\bibinfo {chapter} {2}\BibitemShut {NoStop}%
\bibitem [{\citenamefont {Domb}\ and\ \citenamefont
  {Lebowitz}(1986)}]{DombLebowitz:10}%
  \BibitemOpen
  \bibinfo {editor} {\bibnamefont {Domb}, \bibfnamefont {C.}}, \ and\ \bibinfo
  {editor} {\bibfnamefont {J.~L.}\ \bibnamefont {Lebowitz}},\ Eds. (\bibinfo
  {year} {1986}),\ \href@noop {} {\emph {\bibinfo {title} {Phase Transitions
  and Critical Phenomena}}},\ Vol.~\bibinfo {volume} {10}\ (\bibinfo
  {publisher} {Academic Press},\ \bibinfo {address} {London})\BibitemShut
  {NoStop}%
\bibitem [{\citenamefont {Einhorn}\ and\ \citenamefont
  {Jones}(2007)}]{1126-6708-2007-04-051}%
  \BibitemOpen
  \bibfield  {author} {\bibinfo {author} {\bibnamefont {Einhorn}, \bibfnamefont
  {M.~B.}}, \ and\ \bibinfo {author} {\bibfnamefont {D.~T.}\ \bibnamefont
  {Jones}}} (\bibinfo {year} {2007}),\ \href@noop {} {\bibfield  {journal}
  {\bibinfo  {journal} {Journal of High Energy Physics}\ }\textbf {\bibinfo
  {volume} {2007}}~(\bibinfo {number} {04}),\ \bibinfo {pages}
  {051}}\BibitemShut {NoStop}%
\bibitem [{\citenamefont {Ioffe}(1994)}]{Ioffe1994}%
  \BibitemOpen
  \bibfield  {author} {\bibinfo {author} {\bibnamefont {Ioffe}, \bibfnamefont
  {D.}}} (\bibinfo {year} {1994}),\ \href {\doibase 10.1007/BF02186818}
  {\bibfield  {journal} {\bibinfo  {journal} {J. Stat. Phys.}\ }\textbf
  {\bibinfo {volume} {74}}~(\bibinfo {number} {1}),\ \bibinfo {pages}
  {411}}\BibitemShut {NoStop}%
\bibitem [{\citenamefont {Jasnow}(1986)}]{Jasnow:domb-v10}%
  \BibitemOpen
  \bibfield  {author} {\bibinfo {author} {\bibnamefont {Jasnow}, \bibfnamefont
  {D.}}} (\bibinfo {year} {1986}),\ in\  \cite{DombLebowitz:10},\
  Chap.~\bibinfo {chapter} {3}\BibitemShut {NoStop}%
\bibitem [{\citenamefont {Lalo\"e}(2001)}]{1.1356698}%
  \BibitemOpen
  \bibfield  {author} {\bibinfo {author} {\bibnamefont {Lalo\"e}, \bibfnamefont
  {F.}}} (\bibinfo {year} {2001}),\ \href {\doibase
  http://dx.doi.org/10.1119/1.1356698} {\bibfield  {journal} {\bibinfo
  {journal} {Am. J. Phys.}\ }\textbf {\bibinfo {volume} {69}}~(\bibinfo
  {number} {6}),\ \bibinfo {pages} {655}}\BibitemShut {NoStop}%
\bibitem [{\citenamefont {Low}\ and\ \citenamefont
  {Manohar}(2002)}]{PhysRevLett.88.101602}%
  \BibitemOpen
  \bibfield  {author} {\bibinfo {author} {\bibnamefont {Low}, \bibfnamefont
  {I.}}, \ and\ \bibinfo {author} {\bibfnamefont {A.~V.}\ \bibnamefont
  {Manohar}}} (\bibinfo {year} {2002}),\ \href {\doibase
  10.1103/PhysRevLett.88.101602} {\bibfield  {journal} {\bibinfo  {journal}
  {Phys. Rev. Lett.}\ }\textbf {\bibinfo {volume} {88}},\ \bibinfo {pages}
  {101602}}\BibitemShut {NoStop}%
\bibitem [{\citenamefont {Miracle-Sole}(2013)}]{Miracle-Sole:2013}%
  \BibitemOpen
  \bibfield  {author} {\bibinfo {author} {\bibnamefont {Miracle-Sole},
  \bibfnamefont {S.}}} (\bibinfo {year} {2013}),\ \href@noop {} {\bibfield
  {journal} {\bibinfo  {journal} {Scholarpedia}\ }\textbf {\bibinfo {volume}
  {8}},\ \bibinfo {pages} {31226}}\BibitemShut {NoStop}%
\bibitem [{\citenamefont {Nogawa}\ \emph {et~al.}(2011)\citenamefont {Nogawa},
  \citenamefont {Ito},\ and\ \citenamefont {Watanabe}}]{Nogawa2011}%
  \BibitemOpen
  \bibfield  {author} {\bibinfo {author} {\bibnamefont {Nogawa}, \bibfnamefont
  {T.}}, \bibinfo {author} {\bibfnamefont {N.}~\bibnamefont {Ito}}, \ and\
  \bibinfo {author} {\bibfnamefont {H.}~\bibnamefont {Watanabe}}} (\bibinfo
  {year} {2011}),\ \href {\doibase 10.1103/PhysRevE.84.061107} {\bibfield
  {journal} {\bibinfo  {journal} {Phys. Rev. E}\ }\textbf {\bibinfo {volume}
  {84}},\ \bibinfo {pages} {061107}}\BibitemShut {NoStop}%
\bibitem [{\citenamefont {Schlosshauer}(2005)}]{RevModPhys.76.1267}%
  \BibitemOpen
  \bibfield  {author} {\bibinfo {author} {\bibnamefont {Schlosshauer},
  \bibfnamefont {M.}}} (\bibinfo {year} {2005}),\ \href {\doibase
  10.1103/RevModPhys.76.1267} {\bibfield  {journal} {\bibinfo  {journal} {Rev.
  Mod. Phys.}\ }\textbf {\bibinfo {volume} {76}},\ \bibinfo {pages}
  {1267}}\BibitemShut {NoStop}%
\bibitem [{\citenamefont {Sethna}(2006)}]{sethna}%
  \BibitemOpen
  \bibfield  {author} {\bibinfo {author} {\bibnamefont {Sethna}, \bibfnamefont
  {J.~P.}}} (\bibinfo {year} {2006}),\ \href@noop {} {\emph {\bibinfo {title}
  {Statistical Mechanics : Entropy, Order Parameters, and Complexity}}},\
  \bibinfo {series} {Oxford Master Series in Statistical, Computational, and
  Theoretical Physics}\ No.~\bibinfo {number} {14}\ (\bibinfo  {publisher}
  {Oxford University Press},\ \bibinfo {address} {Oxford})\BibitemShut
  {NoStop}%
\bibitem [{\citenamefont {Shlosman}(1989)}]{shlosman1989}%
  \BibitemOpen
  \bibfield  {author} {\bibinfo {author} {\bibnamefont {Shlosman},
  \bibfnamefont {S.~B.}}} (\bibinfo {year} {1989}),\ \href@noop {} {\bibfield
  {journal} {\bibinfo  {journal} {Comm. Math. Phys.}\ }\textbf {\bibinfo
  {volume} {125}}~(\bibinfo {number} {1}),\ \bibinfo {pages} {81}}\BibitemShut
  {NoStop}%
\bibitem [{\citenamefont {Symanzik}(1970)}]{symanzik1970}%
  \BibitemOpen
  \bibfield  {author} {\bibinfo {author} {\bibnamefont {Symanzik},
  \bibfnamefont {K.}}} (\bibinfo {year} {1970}),\ \href {\doibase
  10.1007/BF01645494} {\bibfield  {journal} {\bibinfo  {journal} {Comm. Math.
  Phys.}\ }\textbf {\bibinfo {volume} {16}}~(\bibinfo {number} {1}),\ \bibinfo
  {pages} {48}}\BibitemShut {NoStop}%
\bibitem [{\citenamefont {Weinberg}(1996)}]{WeinbergQFT:2}%
  \BibitemOpen
  \bibfield  {author} {\bibinfo {author} {\bibnamefont {Weinberg},
  \bibfnamefont {S.}}} (\bibinfo {year} {1996}),\ \href@noop {} {\emph
  {\bibinfo {title} {The Quantum Theory of Fields}}},\ Vol.~\bibinfo {volume}
  {2}\ (\bibinfo  {publisher} {Cambridge University Press},\ \bibinfo {address}
  {Cambridge})\BibitemShut {NoStop}%
\end{thebibliography}%

\end{document}